\documentclass[twocolumn,prl,amsmath,amssymb,aps,showpacs,superscriptaddress]{revtex4-1}
\usepackage{graphicx}
\usepackage{dcolumn}
\usepackage{bm}
\usepackage{}[multicol]

\newcommand{\eek}{$(e,e^{\prime}K^{+})$}

\newcommand*{\TOHOKU}{Graduate School of Science, Tohoku University, Sendai, Miyagi 980-8578, Japan}
\newcommand*{\KYOTO}{Graduate School of Science, Kyoto University, Kyoto, Kyoto 606-8502, Japan}
\newcommand*{\MAINZ}{Institute for Nuclear Physics, Johannes Gutenberg-University, D-55099 Mainz, Germany}
\newcommand*{\NCarolina}{Department of Physics, North Carolina A$\&$T State University,Greensboro, NC 27411, USA}
\newcommand*{\Hampton}{Department of Physics, Hampton University, Hampton, VA 23668, USA}
\newcommand*{\Zagreb}{Department of Physics $\&$ Department of Applied Physics, University of Zagreb, HR-10000 Zagreb, Croatia}
\newcommand*{\Yervan}{A.I.Alikhanyan National Science Laboratory, Yerevan 0036, Armenia}
\newcommand*{\FIU}{Department of Physics, Florida International University, Miami, FL 27411, USA}
\newcommand*{\Christpher}{Department of Physics, Computer Science $\&$ Engineering, Christopher Newport University, Newport News, VA, USA 23606}
\newcommand*{\JLAB}{Thomas Jefferson National Accelerator Facility (JLab), Newport News, VA 23606, USA}
\newcommand*{\Bari}{Istituto Nazionale di Fisica Nucleare, Sezione di Bari and University of Bari, I-70126 Bari, Italy}
\newcommand*{\Southern}{Department of Physics, Southern University at New Orleans,New Orleans, LA 70126, USA}
\newcommand*{\NCarolinatwo}{Department of Physics, University of North Carolina at Wilmington, Wilmington, NC 28403, USA}
\newcommand*{\Roma}{INFN, Sezione Sanit$\grave{a}$ and Istituto Superiore di Sanit$\grave{a}$, 00161 Rome, Italy}
\newcommand*{\Lanzhou}{Nuclear Physics Institute, Lanzhou University, Gansu 730000, China}
\newcommand*{\Mississippi}{Mississippi State University, Mississippi State, Mississippi 39762, USA}
\newcommand*{\James}{Department of Physics and Astronomy, James Madison University, Harrisonburg, VA 22807, USA}
\newcommand*{\Rico}{Divisi\'{o}n de Ciencias y Tecnolog\'{i}a, Universidad Ana G. M\'{e}ndez, Recinto de Cupey, San Juan 00926, Puerto Rico}
\newcommand*{\VMI}{Department of Physics $\&$ Astronomy, Virginia Military Institute, Lexington, Virginia 24450, USA}
\newcommand*{\Yamagata}{Department of Physics, Yamagata University, Yamagata, 990-8560, Japan}
\newcommand*{\Xavier}{Department of Physics, Xavier University of Louisiana, New Orleans, LA 70125, USA}
\newcommand*{\Hou}{Department of Physics, University of Houston, Houston, Texas 77204, USA}

\begin{document}
\preprint{APS/Li9L}
\title{Spectroscopy of $A=9$ hyperlithium by the {\eek} reaction}

\author{T.~Gogami}
\affiliation{\KYOTO}
\affiliation{\TOHOKU}
\author{C.~Chen}
\affiliation{\Hampton}
\author{D.~Kawama}
\affiliation{\TOHOKU}
\author{P.~Achenbach}
\affiliation{\MAINZ}
\author{A.~Ahmidouch}
\affiliation{\NCarolina}
\author{I.~Albayrak}
\affiliation{\Hampton}
\author{D.~Androic}
\affiliation{\Zagreb}
\author{A.~Asaturyan}
\affiliation{\Yervan}
\author{R.~Asaturyan}\thanks{Deceased}
\affiliation{\Yervan}
\author{O.~Ates}
\affiliation{\Hampton}
\author{P.~Baturin}
\affiliation{\FIU}
\author{R.~Badui}
\affiliation{\FIU}
\author{W.~Boeglin}
\affiliation{\FIU}
\author{J.~Bono}
\affiliation{\FIU}
\author{E.~Brash}
\affiliation{\Christpher}
\author{P.~Carter}
\affiliation{\Christpher}

\author{A.~Chiba}
\affiliation{\TOHOKU}
\author{E.~Christy}
\affiliation{\Hampton}
\author{S.~Danagoulian}
\affiliation{\NCarolina}
\author{R.~De~Leo}
\affiliation{\Bari}
\author{D.~Doi}
\affiliation{\TOHOKU}
\author{M.~Elaasar}
\affiliation{\Southern}
\author{R.~Ent}
\affiliation{\JLAB}
\author{Y.~Fujii}
\affiliation{\TOHOKU}
\author{M.~Fujita}
\affiliation{\TOHOKU}
\author{M.~Furic}
\affiliation{\Zagreb}
\author{M.~Gabrielyan}
\affiliation{\FIU}
\author{L.~Gan}
\affiliation{\NCarolinatwo}
\author{F.~Garibaldi}
\affiliation{\Roma}
\author{D.~Gaskell}
\affiliation{\JLAB}
\author{A.~Gasparian}
\affiliation{\NCarolina}
\author{Y.~Han}
\affiliation{\Hampton}
\author{O.~Hashimoto}\thanks{Deceased}
\affiliation{\TOHOKU}
\author{T.~Horn}
\affiliation{\JLAB}
\author{B.~Hu}
\affiliation{\Lanzhou}
\author{Ed.V.~Hungerford}
\affiliation{\Hou}
\author{M.~Jones}
\affiliation{\JLAB}
\author{H.~Kanda}
\affiliation{\TOHOKU}
\author{M.~Kaneta}
\affiliation{\TOHOKU}
\author{S.~Kato}
\affiliation{\Yamagata}
\author{M.~Kawai}
\affiliation{\TOHOKU}

\author{H.~Khanal}
\affiliation{\FIU}
\author{M.~Kohl}
\affiliation{\Hampton}
\author{A.~Liyanage}
\affiliation{\Hampton}
\author{W.~Luo}
\affiliation{\Lanzhou}
\author{K.~Maeda}
\affiliation{\TOHOKU}
\author{A.~Margaryan}
\affiliation{\Yervan}
\author{P.~Markowitz}
\affiliation{\FIU}
\author{T.~Maruta}
\affiliation{\TOHOKU}
\author{A.~Matsumura}
\affiliation{\TOHOKU}
\author{V.~Maxwell}
\affiliation{\FIU}
\author{D.~Meekins}
\affiliation{\JLAB}
\author{A.~Mkrtchyan}
\affiliation{\Yervan}
\author{H.~Mkrtchyan}
\affiliation{\Yervan}
\author{S.~Nagao}
\affiliation{\TOHOKU}
\author{S.N.~Nakamura}
\affiliation{\TOHOKU}
\author{A.~Narayan}
\affiliation{\Mississippi}
\author{C.~Neville}
\affiliation{\FIU}
\author{G.~Niculescu}
\affiliation{\James}
\author{M.I.~Niculescu}
\affiliation{\James}
\author{A.~Nunez}
\affiliation{\FIU}
\author{Nuruzzaman}
\affiliation{\Mississippi}
\author{Y.~Okayasu}
\affiliation{\TOHOKU}
\author{T.~Petkovic}
\affiliation{\Zagreb}
\author{J.~Pochodzalla}
\affiliation{\MAINZ}
\author{X.~Qiu}
\affiliation{\Lanzhou}
\author{J.~Reinhold}
\affiliation{\FIU}
\author{V.M.~Rodriguez}
\affiliation{\Rico}
\author{C.~Samanta}
\affiliation{\VMI}
\author{B.~Sawatzky}
\affiliation{\JLAB}
\author{T.~Seva}
\affiliation{\Zagreb}
\author{A.~Shichijo}
\affiliation{\TOHOKU}
\author{V.~Tadevosyan}
\affiliation{\Yervan}
\author{L.~Tang}
\affiliation{\Hampton}
\affiliation{\JLAB}
\author{N.~Taniya}
\affiliation{\TOHOKU}
\author{K.~Tsukada}
\affiliation{\TOHOKU}
\author{M.~Veilleux}
\affiliation{\Christpher}
\author{W.~Vulcan}
\affiliation{\JLAB}
\author{F.R.~Wesselmann}
\affiliation{\Xavier}
\author{S.A.~Wood}
\affiliation{\JLAB}
\author{T.~Yamamoto}
\affiliation{\TOHOKU}
\author{L.~Ya}
\affiliation{\Hampton}
\author{Z.~Ye}
\affiliation{\Hampton}
\author{K.~Yokota}
\affiliation{\TOHOKU}
\author{L.~Yuan}
\affiliation{\Hampton}
\author{S.~Zhamkochyan}
\affiliation{\Yervan}
\author{L.~Zhu}
\affiliation{\Hampton}

\collaboration{ HKS (JLab E05-115) Collaboration }

\date{\today}

\begin{abstract}
  Missing mass spectroscopy with the {\eek} reaction was
  performed at Jefferson Laboratory's Hall C for the
  neutron rich $\Lambda$ hypernucleus $^{9}_{\Lambda}{\rm Li}$.
  The ground-state (g.s.) energy was obtained to be
  $B_{\Lambda}^{\rm g.s.}=8.84\pm0.17^{\rm stat.}\pm0.15^{\rm sys.}~{\rm MeV}$
  by using shell model calculations of a cross section ratio
  and an energy separation of the spin doublet states ($3/2^{+}_1$ and $5/2^{+}_1$).
  In addition, peaks that are considered to be
  states of [$^{8}{\rm Li}(3^{+})\otimes s_{\Lambda}=3/2^{+}_{2}, 1/2^{+}$]
  and [$^{8}{\rm Li}(1^{+})\otimes s_{\Lambda}=5/2^{+}_{2}, 7/2^{+}$]
  were observed at $E_{\Lambda}({\rm no.~2})=1.74\pm0.27^{\rm stat.}\pm0.11^{\rm sys.}~{\rm MeV}$ and
  $E_{\Lambda}({\rm no.~3})=3.30\pm0.24^{\rm stat.}\pm0.11^{\rm sys.}~{\rm MeV}$, respectively.
  The $E_{\Lambda}({\rm no.~3})$ is larger than shell model predictions
  by a few hundred keV, and the difference would indicate
  that a ${\rm ^{5}He}+t$ structure is more developed
  for the $3^{+}$ state than those for
  the $2^{+}$ and $1^{+}$ states in a core nucleus $^{8}{\rm Li}$ as a cluster model calculation suggests.
\end{abstract}
\maketitle
%
The nucleon-nucleon interaction ($NN$) is well understood
thanks to the
rich data set from scattering and nuclear spectroscopy experiments.
On the other hand, hyperon-nucleon ($YN$)
and hyperon-hyperon ($YY$) interactions
are less understood because experimental data
for the strangeness sector are scarce.
Scattering experiments are difficult for
hyperons due to their short lifetimes.
Data from hyperon scattering experiments are still limited~\cite{cite:alex},
although a $\Sigma$-proton scattering experiment
was recently carried out at J-PARC~\cite{cite:e40}.
Therefore, hypernuclear spectroscopy
plays a vital role in investigations of $YN$ and $YY$ interactions.

The $\Lambda N \mathchar`- \Sigma N$ coupling is one of the important effects
in the $\Lambda N$ interaction.
The energy difference between $^{4}_{\Lambda}$H and $^{4}_{\Lambda}$He
is firm evidence of the charge symmetry breaking (CSB)
in the $\Lambda N$ interaction~\cite{cite:esser,cite:florian,cite:yamasan},
and the $\Lambda N \mathchar`- \Sigma N$ coupling
is considered to be key to solving the $\Lambda N$ CSB issue~\cite{cite:gibson,cite:akaishi,cite:nogga}.
A neutron rich system is a good environment
in which to investigate the $\Lambda N \mathchar`- \Sigma N$ coupling
because it is predicted that
the $\Sigma$ mixing probability in a neutron rich system
is rather higher and that the energy structure
is more affected by the coupling
compared to so called normal $\Lambda$ hypernuclei~\cite{cite:umeya}.
However, there are few data on neutron rich $\Lambda$ hypernuclei.
For example, superheavy hyperhydrogen $^{6}_{\Lambda}$H
and superheavy hyperlithium $^{10}_{\Lambda}$Li
were investigated via double charge exchange reactions using hadron beams.
The FINUDA Collaboration identified three events
that are interpreted as $^{6}_{\Lambda}$H~\cite{cite:finuda}.
Experiments at J-PARC and KEK, on the other hand, 
were not able to determine the $\Lambda$ binding energies
of $^{6}_{\Lambda}$H~\cite{cite:sugimura,cite:honda}
and $^{10}_{\Lambda}$Li~\cite{cite:saha}
due to either low statistics or insufficient energy resolution.
In this Letter, we report new spectroscopic data
of a neutron rich $\Lambda$ hypernucleus $^{9}_{\Lambda}$Li
for which we performed missing mass spectroscopy
with the {\eek} reaction at Jefferson Laboratory's (JLab) experimental Hall C.

A difference of $\Lambda$ binding energies between
mirror hypernuclei is a benchmark of CSB in the $\Lambda N$ interaction.
$\Lambda N$ CSB was discussed
in $s$-shell hypernuclei~\cite{cite:dalitz,cite:bodmer,cite:yamasan,cite:gazda}, 
and the interest is extended
to CSB in $p$-shell hypernuclear systems~\cite{cite:hiyama,cite:gal,cite:finuda2}.
We present new binding energy data
for $^{9}_{\Lambda}$Li which are compared with
that of the mirror hypernucleus $^{9}_{\Lambda}{\rm B}$.

We performed a series of $\Lambda$ binding energy measurements
for several $p$-shell hypernuclei with
a new magnetic spectrometer system HKS-HES
(Experiment JLab E05-115)~\cite{cite:hes},
and results for $^{7}_{\Lambda}$He~\cite{cite:7LHe},
$^{10}_{\Lambda}$Be~\cite{cite:10LBe} and 
$^{12}_{\Lambda}$B~\cite{cite:12LB} were published.
We also took data with a $^{9}$Be target to produce $^{9}_{\Lambda}$Li.
A continuous $E_{e}=2.344$-GeV electron beam
was impinged on a 188-mg/cm$^{2}$ $^{9}$Be target.
The beam had a typical intensity on target 
of about $38~\mu$A with a beam bunch cycle of 2~ns.
A total of 5.3~C ($=3.3\times10^{19}$ electrons) was delivered to the target.
The scattered electron and $K^{+}$
with central momenta of $p_{e^{\prime}}=0.844$ and
$p_{K}=1.200$~GeV/$c$ were measured by the HES and HKS~\cite{cite:fujii}, respectively.
The HES and HKS spectrometers have momentum
resolutions of $\Delta p/p \simeq 2 \times 10^{-4}$ FWHM
allowing us to achieve the best energy resolution
in missing mass spectroscopy of hypernuclei~\cite{cite:12LB}.

In order to calibrate the absolute energy
in the missing mass spectrum,
we used the reactions $p${\eek}$\Lambda$
and $p${\eek}$\Sigma^{0}$ on a polyethylene target (CH$_x$)
to produce $\Lambda$ and $\Sigma^{0}$ hyperons
for which we know the masses with uncertainties
of only $\pm 6$ and $\pm 24$~keV, respectively~\cite{cite:pdg}.
The calibration used
the same
spectrometer settings as those
for hypernuclear production
thanks to the large momentum acceptances of the HES and HKS
($\Delta p_{{\rm accept}} /p_{{\rm central}} = $ $\pm 17.5\%$
and $\pm 12.5\%$, respectively),
minimizing the systematic error on the binding energy measurement.
The systematic error was evaluated
by a Geant4 Monte Carlo (MC) simulation~\cite{cite:geant4_1,cite:geant4_2}
in which precise geometry, materials, and magnetic fields
were modeled.
The calibration analysis that
was used for the real data 
was applied to several sets of dummy data
from the MC simulation to estimate the systematic error
on the binding energy.
As a result, the systematic errors
originating from the energy calibration
for the $\Lambda$ binding energy and the excitation energy
were evaluated to be $\Delta B_{\Lambda}^{{\rm sys.}}=0.11$
and $\Delta E_{\Lambda}^{{\rm sys.}}=0.05$ ~MeV, respectively.
Refer to Refs.~\cite{cite:hes, cite:12LB} for details
about the calibration method.

In the hadron arm of the HKS spectrometer,
backgrounds of $\pi^{+}$'s and protons
were rejected to identify $K^{+}$'s
both on-line (data taking trigger) and off-line (data analysis).
To reduce the trigger rate to less than $2~{\rm kHz}$, allowing
a data acquisition live time of over  $90\%$,
we incorporated two types of Cherenkov detectors
(AC and WC; radiation media of a hydrophobic aerogel and a deionized water with
refractive indices of $n=1.05$ and 1.33, respectively) in the trigger. 
Off-line, the $K^{+}$ identification (KID) was performed by the following three criteria:
(KID-1)~coincidence time analysis, 
(KID-2)~light yield analysis in AC and WC,
(KID-3)~analysis of particle squared mass.
The coincidence time is defined as $t_{\rm coin}=t_{e^{\prime}}-t_{K}$
where $t_{e^{\prime},K}$ are the times at target.
The $t_{e',K}=t^{\rm TOF}-\Bigl(\frac{l}{v_{e^{\prime},K}}\Bigr)$ were calculated event by event
by using the velocity $v_{e^{\prime},K}$,
the time at the time-of-flight (TOF) detector $t^{\rm TOF}$,
and the path length ($l$)
from the target to the TOF detector for each particle.
The velocity $v_{e^{\prime},K}$ was obtained from
the particle momentum which was calculated by the backward transfer matrix
with assumptions of the masses of $e^{\prime}$ and  $K^{+}$ for
particles in HES and HKS, respectively.
A coincidence event of ($e^{\prime}$ - $K^{+}$) could be
identified with a resolution of 0.64-ns (FWHM) in the
coincidence time. 
Peaks of other coincidence reactions such as ($e^{\prime}$ - $\pi^{+}$) and ($e^{\prime}$ - $p$)
are located at different times with
respect to the ($e^{\prime}$ - $K^{+}$) one
because of the wrong assumptions of particle masses for $\pi^{+}$'s and protons.
The other coincidence events
and most of the accidental coincidence events could be removed
by a coincidence time selection 
with a time gate of $\pm 1$-ns width for the real ($e^{\prime}$ - $K^{+}$) coincidence peak~\cite{cite:hes}.
Only $0.047\%$ and $0.019\%$ of the $\pi^{+}$'s and protons, respectively,
survived when KID-2 and 3 were used, whereas $>80\%$ of the $K^{+}$'s remained
after these cuts~\cite{cite:bucking}.

\begin{figure}[!htbp]
  \includegraphics[width=8cm]{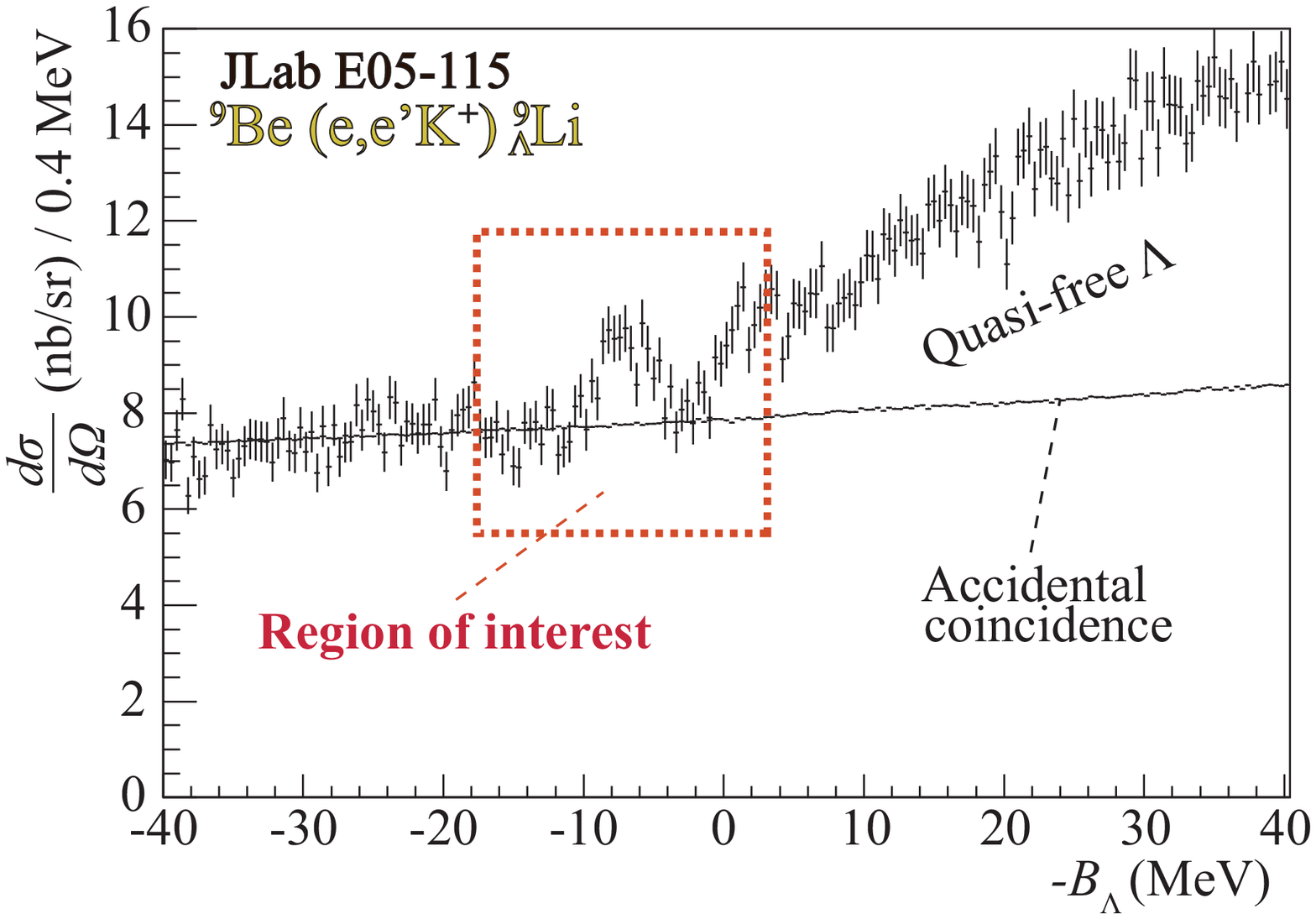}
  \caption{Differential cross section of the $^{9}$Be{\eek}$^{9}_{\Lambda}$Li reaction
    as a function of $-B_{\Lambda}$.
    Events exceeding over the accidental coincidence background in the
    bound region ($-B_{\Lambda}<0$) were analyzed in the present Letter.}
  \label{fig:cs}
\end{figure}

\begin{figure}[!htbp]
  \includegraphics[width=8cm]{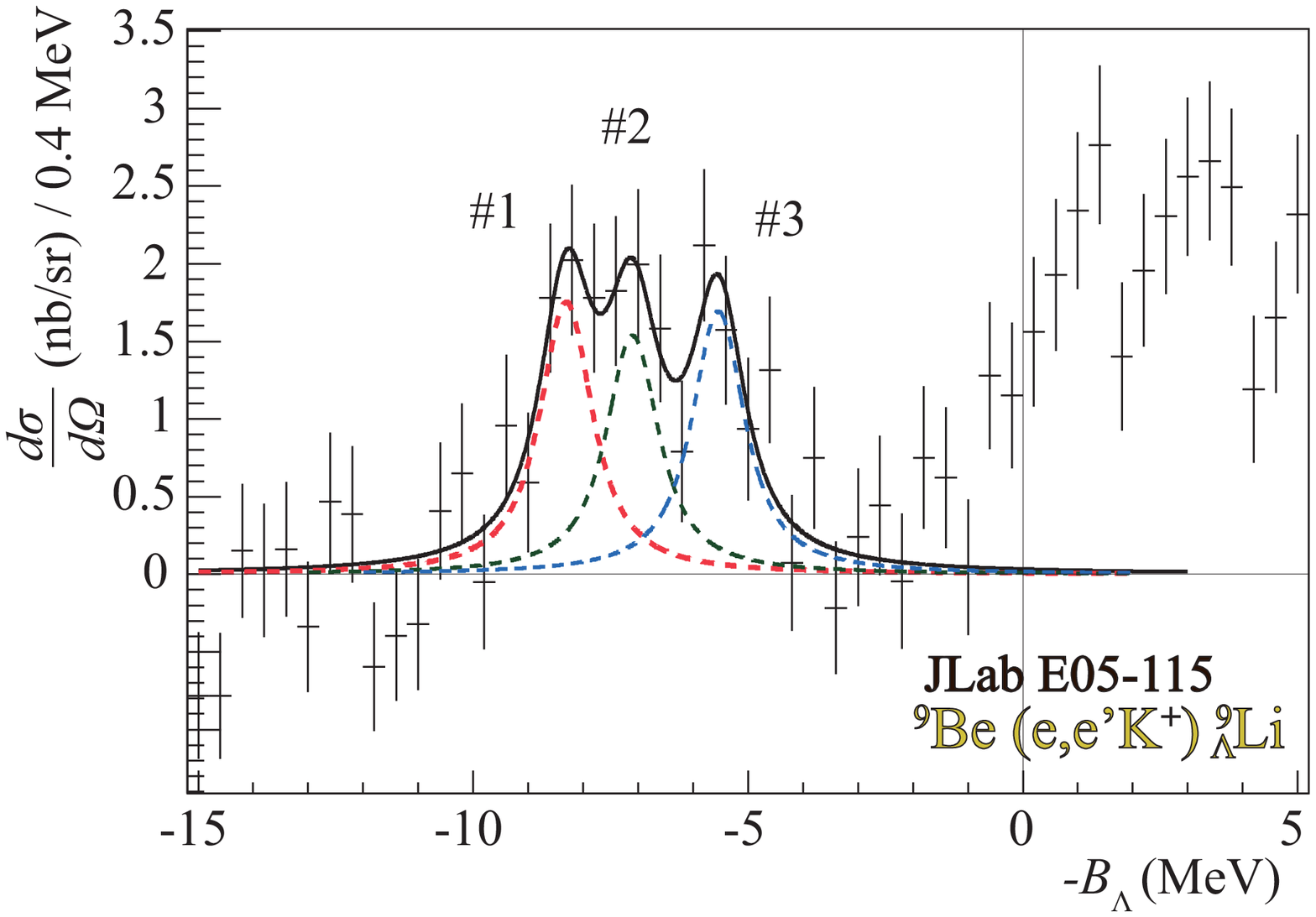}
  \caption{Fit of the $^{9}$Be{\eek}$^{9}_{\Lambda}$Li spectrum by three Voigt functions
    after the accidental coincidence events
    obtained by the mixed event analysis (Fig.~\ref{fig:cs}) was subtracted.}
  \label{fig:fit}
\end{figure}

Figure~\ref{fig:cs} shows the differential cross section
as a function of $-B_{\Lambda}$ for the reaction of $^{9}$Be{\eek}$^{9}_{\Lambda}$Li.
The abscissa is $-B_{\Lambda} = -[M(^{8}{\rm Li})+M_{\Lambda}-M_{\rm HYP}]$
where $M(^{8}{\rm Li})$ and $M_{\Lambda}$ 
are the masses of the $^{8}{\rm Li}$ core nucleus 
and the $\Lambda$ which
are 7471.36~MeV/$c^{2}$~\cite{cite:wang} and 1115.68~MeV/$c^{2}$~\cite{cite:pdg},
respectively.
The mass of $M(^{9}{\rm Be})=8392.75~{\rm MeV}/c^{2}$~\cite{cite:wang} was used for
the target nucleus $^{9}{\rm Be}$ to calculate $M_{\rm HYP}$.
The ordinate is the differential cross section
in the laboratory frame
for the $(\gamma^{*},K^{+})$ reaction $\Bigl(\frac{d \sigma}{d \Omega_{K}} \Bigr) \Bigr|_{{\rm HKS}}$
that is described in Refs.~\cite{cite:7LHe,cite:10LBe}.
It must be noted that $Q^{2}$ ($=-q^{2}$ where $q$ is the four-momentum
transfer to a virtual photon) was small [$Q^{2} = 0.01$~(GeV$/c)^{2}$]
in our experimental setup, and thus,
the virtual photon may be treated as almost a real photon.
The $K^{+}$ scattering angle with respect to the virtual photon 
was $\theta^{\rm laboratory}_{\gamma K}\simeq7~{\rm deg}$.
As the electron spectrometer was tilted out of the horizontal plane~\cite{cite:hes}, 
the angle between the electron scattering plane and the reaction plane $\phi_{K}$ 
was approximately $90^{\circ}$.
The distribution of accidental coincidence events
shown in Fig.~\ref{fig:cs}
was obtained by the mixed event analysis
in which the missing mass
was reconstructed with random combinations of $e^{\prime}$ and $K^{+}$ from different events~\cite{cite:7LHe_0}.
The accidental background distribution was subtracted
as shown in Fig.~\ref{fig:fit}, 
and residual events in a region of $-B_{\Lambda}<0$ were analyzed
as bound states of $^{9}_{\Lambda}$Li.
Three doublet states 
for which a $\Lambda$ residing in the $s$ orbit
couples with the $2^{+}$ (ground state), $1^{+}$, and $3^{+}$ states
of the core nucleus $^{8}$Li
are expected to be largely populated
in the $^{9}_{\Lambda}$Li spectrum~\cite{cite:sotona, cite:motoba}.
In addition, the energy spacings between the states in each spin doublet
are theoretically expected to
be at most about 0.6~MeV  making them difficult to separate
given the expected experimental resolution.
Therefore, we used three Voigt functions with
the same width for fitting the cross section spectrum.
The fitting result with $\chi^{2}/{\rm n.d.f.}=22.24/22$ is summarized in Table~\ref{tab:result}.
The full width at half maximum of the Voigt function for each peak
was found to be $1.1\pm0.4$~MeV which is
consistent with that expected in the MC simulation.
The cross-section ratios of peaks no.~2 and no.~3 to
that of peak no.~1 are $0.88\pm0.13$ and $0.96\pm0.15$, respectively, 
whereas the ratios of the corresponding spectroscopic factors $C^{2}S$ are 0.60 and 0.65, respectively,
as measured in the $^{9}{\rm Be}(t,\alpha)^{8}{\rm Li}$ reaction~\cite{cite:liu}.
\begin{table*}
  \caption{Fitting result of the $^{9}{\rm Be}${\eek}$^{9}_{\Lambda}{\rm Li}$ spectrum in JLab E05-115.
    Three Voigt functions were used for the fitting.
    The $\Lambda$ binding energy of the ground state $B_{\Lambda}^{\rm g.s.}$
    and the excitation energy $E_{\Lambda}$ were evaluated
    with the assumption that the cross section ratio of
    the first excited state $5/2^{+}_{1}$ to that of the ground state $3/2^{+}_{1}$
    is 5--7 and the doublet separation is 0.5--0.7~MeV~\cite{cite:umeya,cite:sotona,cite:millener}.}
  \label{tab:result}
  \begin{tabular}{|c|c|c|c|c|}
    \hline
    Peak ID & Possible states & $B_{\Lambda}$ (MeV) & $E_{\Lambda}$ (MeV) & $\Bigl( \frac{d \sigma}{d \Omega_{K}} \Bigr)\Bigr|_{{\rm HKS}}$ (nb/sr) \\ \hline 
    No.~1 & $^{8}{\rm Li} (2^{+}) \otimes s_{\Lambda}$ & $8.31 \pm 0.17 \pm 0.11^{\rm sys.} $ & [$\Delta B_{\Lambda} ({\rm g.s.-} {\rm no.~1}) = 0.53 \pm 0.10^{\rm sys.}$] & $7.6 \pm 0.8^{\rm stat.} \pm 0.8^{\rm sys.}$ \\ 
           & $=3/2^{+}_{1}, 5/2^{+}_{1}$  & ($B_{\Lambda}^{{\rm g.s.}} = 8.84\pm0.17^{{\rm stat.}}\pm0.15^{\rm sys.}$) &   &   \\ \hline 
    No.~2 & $^{8}{\rm Li} (1^{+}) \otimes s_{\Lambda}$ & $7.10 \pm 0.21 \pm 0.11^{\rm sys.} $ & $1.74 \pm 0.27^{\rm stat.} \pm 0.11^{\rm sys.} $ & $6.7 \pm 0.7^{\rm stat.} \pm 0.7^{\rm sys.}$ \\
    & $=3/2^{+}_{2}, 1/2^{+}$ & & & \\ \hline 
    No.~3 & $^{8}{\rm Li} (3^{+}) \otimes s_{\Lambda}$ & $5.54 \pm 0.17 \pm 0.11^{\rm sys.} $ & $3.30 \pm 0.24^{\rm stat.} \pm 0.11^{\rm sys.} $ & $7.3 \pm 0.8^{\rm stat.} \pm 0.7^{\rm sys.}$ \\ 
    & $=5/2^{+}_{2}, 7/2^{+}$ & & & \\ \hline
  \end{tabular}
\end{table*}
Peak no.~1 is considered to be 
the first doublet state, $^{8}{\rm Li}(2^{+};{\rm g.s.}) \otimes s_{\Lambda}=3/2^{+}_{1}, 5/2^{+}_{1}$.
It is predicted that
the production cross section of the $5/2^{+}_{1}$ state is 
larger than that of the ground state $3/2^{+}_{1}$ by a factor of 5--7
and the doublet separation is 0.5--0.7~MeV~\cite{cite:umeya,cite:sotona,cite:millener}.
Assuming this cross section ratio and doublet separation, 
the ground state binding energy is evaluated to be 
greater than the mean value of peak no.~1 by $0.53 \pm 0.10~$MeV [$=\Delta B_{\Lambda}({\rm g.s.-}{\rm no.~1})$]
by a simple simulation
leading to the ground state energy 
$B_{\Lambda}^{{\rm \rm Hall \mathchar`- C}} (^{9}_{\Lambda}{\rm Li; g.s.})=8.84\pm0.17^{{\rm stat.}}\pm0.15^{\rm sys.}$~MeV.
The obtained $B_{\Lambda}$ agrees with
$B_{\Lambda}^{\rm emul.} (^{9}_{\Lambda}{\rm Li; g.s.}) =8.50\pm0.12~{\rm MeV}$~\cite{cite:pniewski},
the mean binding energy of 13 emulsion events,
and 
$B_{\Lambda}^{\rm Hall \mathchar`- A} (^{9}_{\Lambda}{\rm Li; g.s.}) =8.36\pm0.08^{\rm stat.}\pm0.08^{\rm sys.}~{\rm MeV}$~\cite{cite:guido, cite:franco}
within $\pm 2 \sigma$ of the uncertainty.
The weighted average of the above three measurements including our result
is found to be $B_{\Lambda}^{\rm mean} (^{9}_{\Lambda}{\rm Li; g.s.})= 8.47\pm0.08^{\rm total}~{\rm MeV}$.

\begin{figure}[!htbp]
  \includegraphics[width=8cm]{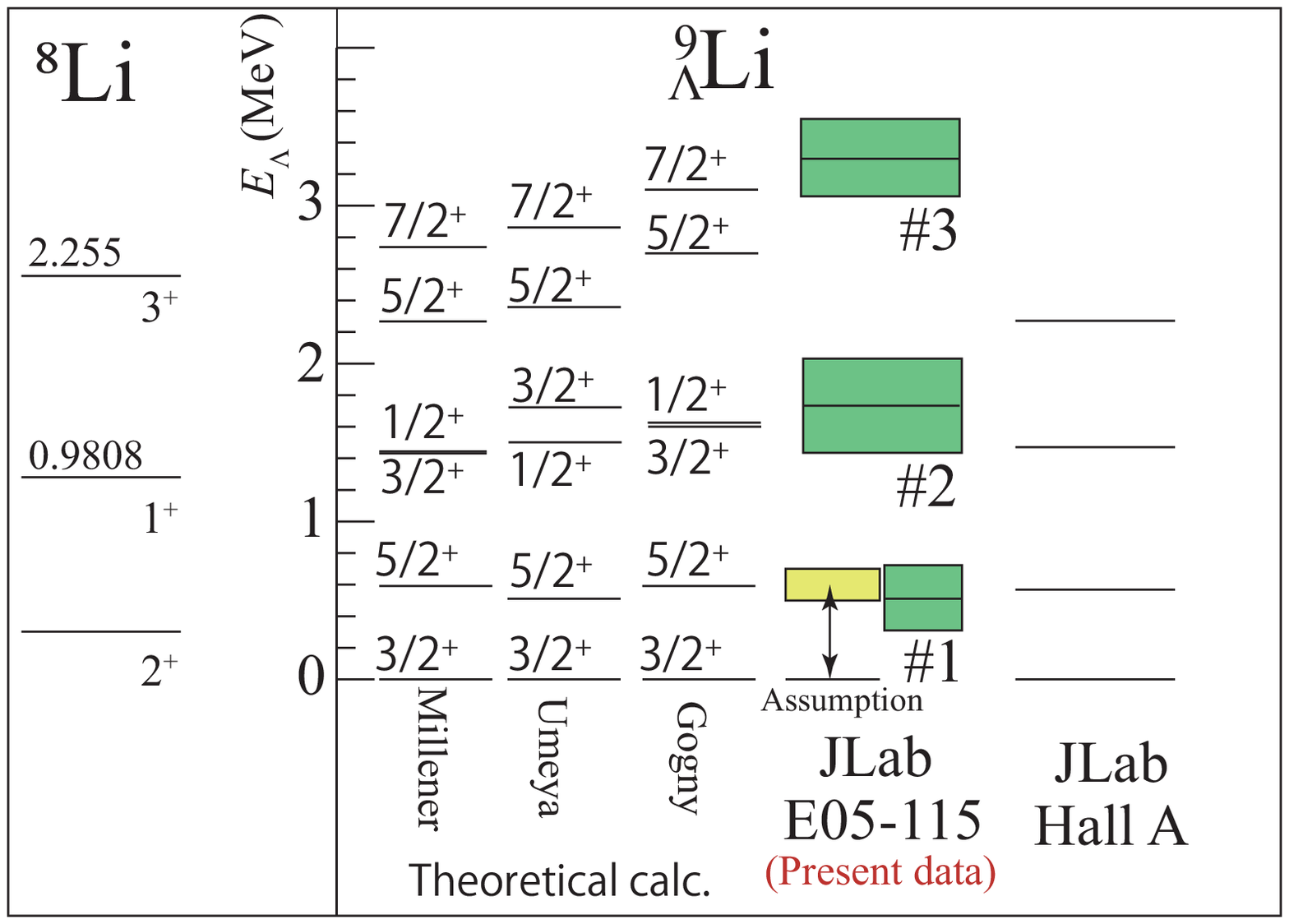}
  \caption{Comparison of the obtained excitation energy
    $E_{\Lambda}$ of $^{9}_{\Lambda}{\rm Li}$ with 
    theoretical calculations~\cite{cite:umeya,cite:millener,cite:gogny}
    and experimental data taken at JLab Hall A~\cite{cite:guido, cite:franco}.
    $E_{\Lambda}$ was obtained with the assumption that
    the cross section ratio of the $5/2^{+}$ state to that of
    the ground state $3/2^{+}$ is 5--7 and
    the doublet separation is 0.5--0.7~MeV~\cite{cite:umeya,cite:sotona,cite:millener}. }
  \label{fig:comp}
\end{figure}

The excitation energies ($E_{\Lambda}$) for peaks no.~2 and no.~3 
were calculated based on the obtained ground state energy $B_{\Lambda}^{{\rm \rm Hall \mathchar`- C}} (^{9}_{\Lambda}{\rm Li; g.s.})$ 
and are shown in Table~\ref{tab:result}.
Figure~\ref{fig:comp} shows a comparison of the obtained
$E_{\Lambda}$ with those of
shell model predictions~\cite{cite:umeya,cite:millener,cite:gogny}
and the experimental data from JLab Hall A~\cite{cite:guido, cite:franco}.
Experimental energy levels of the core nucleus $^{8}$Li taken
from Ref.~\cite{cite:tilley} are shown as well.
The excitation energy of
$E_{\Lambda} ({\rm no.~2})=1.74\pm0.27^{\rm stat.}\pm0.11^{\rm sys.}~{\rm MeV}$ is consistent with
those of the theoretical predictions of $3/2^{+}_2$ and $1/2^{+}$
and the experimental result of JLab Hall A.
For the third doublet which is considered to correspond to peak no.~3
the cross section of the $7/2^{+}$ is
predicted to be larger than that of $5/2^{+}_2$
by a factor of 2 or 3~\cite{cite:sotona,cite:millener},
and thus peak no.~3 is expected to be dominated by the $7/2^{+}$ state.
The energy of peak no.~3
was found to be $E_{\Lambda} ({\rm no.~3})= 3.30\pm0.24^{\rm stat.}\pm0.11^{\rm sys.}~{\rm MeV}$.
It is found that $E_{\Lambda} ({\rm no.~3})$ is larger than the predicted energy of $7/2^{+}$ by a few hundred keV.
$E_{\Lambda}$ could be larger if the core nucleus is deformed
due to a development of clusters
because a spatial overlap between the core nucleus and the $\Lambda$ gets smaller~\cite{cite:isaka}.
A cluster model calculation suggests that a ${{\rm He}^{5}}+t$ structure
is more developed for the $3^{+}$ state than for the $2^{+}$ and $1^{+}$ states
in  $^{8}{\rm Li}$~\cite{cite:stowe}.
The larger energy compared to the shell model predictions
for peak no.~3 may indicate the development of clusters
for the $3^{+}$ state of the core nucleus $^{8}{\rm Li}$.

The highest excitation energy peak
observed by the experiment at JLab Hall A was at $2.27\pm0.09~{\rm MeV}$~\cite{cite:guido, cite:franco}
that differs from $E_{\Lambda}({\rm no.~3})$ by about 1~MeV.
If we assume 0.23~MeV of the energy separation
between the first doublet states 
instead of the assumption of 0.5--0.7~MeV separation, 
the central value of the ground state energy becomes
consistent with that of the emulsion experiment ($B_{\Lambda}^{\rm emul.}$).
Accordingly, the excitation energies are reduced by 0.34~MeV [$=0.53-(8.50-8.31)~{\rm MeV}$]
from those shown in Table~\ref{tab:result} and Fig.~\ref{fig:comp},
and $E_{\Lambda}$(nos.~2 and 3) become more consistent with the theoretical predictions.
However, $E_{\Lambda}({\rm no.3})$ obtained with this different assumption is still far from
the energy of the most excited state observed at JLab Hall A.
Peaks that originate from different states might be observed 
due to a difference in kinematics, such as $Q^{2}$ and
the $K^{+}$-scattering angle with respect to the virtual photon.
However, the relative strength of the cross section for each state in the present experiment 
is predicted not to differ so much from that of JLab Hall A
in DWIA calculations~\cite{cite:bydzovsky} in which elementary amplitudes of
the Saclay-Lyon and BS3 models~\cite{cite:dalibor} are used.
Further studies are necessary to consistently understand these experimental spectra.

Three events of $^{9}_{\Lambda}{\rm B}$ were identified in the emulsion experiment,
and the mean value was reported to be
$B_{\Lambda} (^{9}_{\Lambda}{\rm B; g.s.})=8.29\pm0.18~{\rm MeV}$~\cite{cite:pniewski}.
The difference of $\Lambda$ binding energies between
the $A=9$ isotriplet hypernuclei was found to be
$B_{\Lambda}(^{9}_{\Lambda}{\rm B; g.s.})-B^{{\rm Hall \mathchar`- C}}_{\Lambda}(^{9}_{\Lambda}{\rm Li; g.s.})=-0.55\pm 0.29~{\rm MeV}$
to be compared with the prediction of $-0.054~{\rm MeV}$~\cite{cite:gal}.
There might be an unexpectedly large CSB effect in the $A=9$ isotriplet hypernuclei.
However, the current experimental precision
is not sufficient for $^{9}_{\Lambda}{\rm Li}$ as well as $^{9}_{\Lambda}{\rm B}$
to discuss the $\Lambda N$ CSB in the system.
In order to precisely determine the ground state energy by an experiment
with the {\eek} reaction, the first doublet states would need to be resolved.
The doublet separation of $^{9}_{\Lambda}{\rm Li}$ (between $3/2^{+}$ and $5/2^{+}$ states)
is predicted to be 0.5--0.7~MeV which
is much larger than for other $p$-shell hypernuclei
(e.g. the separation between $1^{-}$ (g.s.) and $2^{-}$ states of
$^{12}_{\Lambda}{\rm C}$ was measured to be
$0.1615\pm0.0003~{\rm MeV}$ ~\cite{cite:kenji}).
This is partially due to a large contribution of the $\Lambda N$-$\Sigma N$ coupling~\cite{cite:umeya}.
Therefore,
an {\eek} experiment with an energy resolution of 0.5~MeV (FWHM) or better 
would be a promising way to precisely determine the ground state energy of $^{9}_{\Lambda}{\rm Li}$.

To summarize, we measured $^{9}_{\Lambda}{\rm Li}$
by missing mass spectroscopy with the {\eek} reaction at JLab Hall C.
We observed three peaks (nos.~1--3) that are considered
to be $s_{\Lambda}$ states coupling with
a $^{^8}{\rm Li}$ nucleus in the $2^{+}$, $1^{+}$, and $3^{+}$ states.
Peak no.~1 that is expected to be
the spin doublet state of $[^{8}{\rm Li}(2^{+})\otimes s_{\Lambda}(=3/2_{1}^{+},5/2_{1}^{+})]$ was analyzed to
obtain the ground state energy.
The ground state energy was determined to
be $B_{\Lambda}^{{\rm Hall \mathchar`- C}}(^{9}_{\Lambda}{\rm Li; g.s.}) = 8.84\pm0.17^{\rm stat.}\pm0.15^{\rm sys.}~{\rm MeV}$
using the assumptions that
the cross section ratio of the first excited state ($5/2^{+}_{1}$) to
that of the ground state ($3/2^{+}_{1}$) is 5--7
and that the doublet energy separation is 0.5--0.7~MeV~\cite{cite:umeya,cite:sotona,cite:millener}.
Peaks no.~2 and no.~3
are considered to be $[^{8}{\rm Li}(1^{+})\otimes s_{\Lambda}(=3/2_{2}^{+},1/2^{+})]$
and $[^{8}{\rm Li}(3^{+})\otimes s_{\Lambda}(=5/2_{2}^{+},7/2^{+})]$ states, respectively.
We obtained excitation energies to be
$E_{\Lambda} ({\rm no.~2})=1.74\pm0.27^{\rm stat.}\pm0.11^{\rm sys.}$~MeV
and $E_{\Lambda} ({\rm no.~3})=3.30\pm0.24^{\rm stat.}\pm0.11^{\rm sys.}$~MeV
by using the $B_{\Lambda}^{{\rm Hall \mathchar`- C}}(^{9}_{\Lambda}{\rm Li; g.s.})$.
$E_{\Lambda}({\rm no.~3})$ is
larger than predicted by shell model calculations
for which different $NN$ and $\Lambda N$ interactions are used 
whereas $E_{\Lambda}({\rm no.~2})$ agrees with the theoretical predictions.
The difference of about a few hundred keV supports
the idea a ${\rm ^{5}He} + t$ structure is more developed 
for the $3^{+}$ state than for 
the $2^{+}$ and $1^{{+}}$ states of the $^{8}{\rm Li}$ nucleus,
as a cluster model calculation suggests~\cite{cite:stowe}.

We thank the JLab staff of the physics, accelerator, and
engineering divisions for support of the experiment.
We thank P.~Byd${\rm \check{z}}$ovsk${\rm \acute{y}}$,
E.~Hiyama, M.~Isaka, D.J.~Millener, T.~Motoba, and A.~Umeya
for valuable exchanges for this Letter.
This Letter was partially supported by the Grant-in-Aid for Scientific Research on
Innovative Areas ``Toward new frontiers  Encounter and synergy of
state-of-the-art astronomical detectors and exotic quantum beams.''
The program was supported by JSPS KAKENHI Grants
No.~JP18H05459, No.~18H01219, No.~17H01121,
No.~12002001, No.~15684005, No.~16GS0201, No.~24$\cdot$4123,
JSPS Core-to-Core Program No.~21002, 
JSPS Strategic Young Researcher Overseas Visits
Program for Accelerating Brain Circulation Grant No.~R2201, and 
JSPS and DAAD under the Japan-Germany Research Cooperative Program. 
We acknowledge support by SPIRITS 2020 of Kyoto University, 
the Graduate Program on Physics for the Universe, Tohoku University (GP-PU), 
U.S. Department of Energy Contracts
No.~DE-AC05-84ER40150, No.~DE-AC05-06OR23177, No.~DE-FG02-99ER41065,
No.~DE-FG02-97ER41047, No.~DE-AC02-06CH11357, No.~DE-FG02-00ER41110, and
No.~DE-AC02-98CH10886, and U.S.-NSF Contracts No.~013815 and No.~0758095.

\end{document}